\documentclass[aps]{revtex4}
\usepackage{graphicx,epsfig,wrapfig,color}
\usepackage{amsmath,amssymb,bm}
\usepackage{color}

\definecolor{darkgreen}{rgb}{0,0.65,0}

\newcommand{\be}{\begin{equation}}
\newcommand{\ee}{\end{equation}}
\newcommand{\ba}{\begin{eqnarray}}
\newcommand{\ea}{\end{eqnarray}}
\newcommand{\di}{\!{\rm d}}
\newcommand{\la}{\langle}
\newcommand{\ra}{\rangle}

\newcommand{\bDelta}{ {\bf\Delta}}
\newcommand{\fslash}[1] {{\not\! #1\,}}

\begin{document}
\newcommand*{\UConn}{Departement of Physics, University of Connecticut,
Storrs, CT 06269, U.S.A.}\affiliation{\UConn}
\title{\boldmath
	The dynamic origins of fermionic $D$-terms}
\author{Jonathan Hudson}\affiliation{\UConn}
\author{Peter Schweitzer}\affiliation{\UConn}
\date{November 2017}
\begin{abstract}
The $D$-term is a particle property defined, similarly to the mass and 
spin, through matrix elements of the energy-momentum tensor.
It is currently not known experimentally for any particle, but the 
$D$-term of the nucleon can be inferred from studies of hard-exclusive 
reactions. In this work we show that the $D$-term of a spin $\frac12$ 
fermion is of dynamical origin: it vanishes for a free fermion.
This is in pronounced contrast to the bosonic case where already a 
free spin-0 boson has a non-zero intrinsic $D$-term as shown in an 
accompanying work. 
We illustrate in two simple models how interactions generate the 
$D$-term of a fermion with an internal structure, the nucleon.
All known matter is composed of elementary fermions. This indicates 
the importance to study this interesting particle property in more 
detail, which will provide novel insights especially on the structure 
of the nucleon. 
\end{abstract}
%
%
%
\maketitle


\section{Introduction}
\label{Sec-1:introduction}

The matrix elements of the energy-momentum tensor (EMT) \cite{Pagels}
provide  most basic information: the mass and spin of a particle.
They also define the $D$-term  \cite{Polyakov:1999gs},
a property not known experimentally for any particle. 
The most direct way to probe EMT matrix elements would 
be scattering off gravitons, which is impractical.
However, information on the EMT form factors can be accessed 
through generalized parton distribution functions (GPDs) which 
enter the description of certain hard exclusive reactions
\cite{Muller:1998fv,Ji:1996ek,Radyushkin:1996nd,Collins:1996fb,
Ji:1998pc,Goeke:2001tz,Diehl:2003ny,Belitsky:2005qn,Guidal:2013rya,
Vanderhaeghen:1998uc,Belitsky:2001ns}.
The second Mellin moments of unpolarized GPDs yield EMT form 
factors. This provides not only the key to access information about 
nucleon's spin decomposition \cite{Ji:1996ek}, but also to its
mechanical properties \cite{Polyakov:2002yz}. 
The $D$-term determines the behavior of unpolarized GPDs in the asymptotic 
limit of renormalization scale $\mu\to\infty$ \cite{Goeke:2001tz}. 
Aspects of the relation of the $D$-term to GPDs were also investigated in 
\cite{Teryaev:2001qm}.

Similarly electric form factors providing information on the 
electric charge distribution \cite{Sachs}, the EMT form factors offer
insights on the energy density, orbital angular momentum density, and 
the distribution of internal forces encoded in the stress tensor and
directly related to the $D$-term \cite{Polyakov:2002yz}. 
The EMT densities allow us to gain insights on the particle stability,
and may have interesting practical applications \cite{Eides:2015dtr}.
For a recent review we refer to \cite{Hudson:2016gnq}.

The nucleon $D$-term has been studied in models, lattice QCD, 
and dispersion relations
\cite{Ji:1997gm,Petrov:1998kf,Schweitzer:2002nm,Ossmann:2004bp,
Wakamatsu:2007uc,Goeke:2007fp,Goeke:2007fq,Cebulla:2007ei,Kim:2012ts,
Hagler:2003jd,Gockeler:2003jf,Negele:2004iu,Bratt:2010jn,Pasquini:2014vua}.
$D$-terms have also been investigated in spin-0 
\cite{Novikov:1980fa,Megias:2004uj,Brommel:2005ee,
Guzey:2005ba,Liuti:2005qj,Mai:2012yc,Hudson-PS-I} and in 
higher-spin \cite{Gabdrakhmanov:2012aa,Perevalova:2016dln} systems.
In all cases the $D$-terms were found negative. The nucleon $D$-term
was also studied in chiral perturbation theory which cannot predict 
its value \cite{Chen:2001pv}. 
The fixed poles in virtual Compton amplitudes discussed in the pre-QCD era 
\cite{Cheng:1970vg} might be related to the $D$-term \cite{Brodsky:2008qu}.

With the $D$-term experimentally unknown, theoretical predictions
are of importance. A particularly interesting question is:
what is the $D$-term of a free particle? The purpose of this work is 
to address this question for fermions. 
To illustrate how instructive it is to investigate this question,
one may recall that the free Dirac equation predicts the anomalous 
magnetic moment $g=2$ of a charged point-like fermion, which is 
derived by coupling the free theory to a weak classical magnetic 
background field. 
In principle, the same is implicitly done by defining the EMT through 
coupling the free theory to a classical background gravitational field 
which for Dirac fields yields the symmetric ``Belifante improved'' EMT. 
Interactions alter the value $g=2$; little for electrons and muons in QED,
far more for protons and neutrons in QCD. But in any case, the free theory 
provides a valuable benchmark to which we can compare results from 
theoretical approaches and eventually experiment.

In an accompanying work, this question was studied for the bosonic case: 
free spin-$0$ bosons have an intrinsic $D$-term $D=-1$.
This prediction pertains to free point-like bosons, although interacting 
theories of extended bosons can be constructed where this value is preserved. 
In general, however, interactions affect the value of $D$ 
\cite{Hudson-PS-I}.

In this work we will show that free non-interacting fermions have
no intrinsic $D$-term. This means that, in contrast to bosons,
fermionic $D$-terms are generated by dynamics which is an
unexpected and highly interesting feature. 
We will illustrate in two simple models how interactions
can generate the $D$-term of a fermion.

The outline of this work is as follows.
After introducing the notation in Sec.~\ref{Sec-2:FF-of-EMT-in-general},  
we will compute the EMT form factors for a free spin $\frac12$ particle
in Sec.~\ref{Sec-3:free-case} and show that the $D$-term of a 
non-interacting fermion vanishes, which has implicitly already 
been stated in literature as we became aware after completing 
this part of our work.
In Sec.~\ref{Sec-4:heuristic} we provide a heuristic argument based
on the 3D density formalism to explain why the $D$-term must be
zero for a free pointlike particle for consistency reasons.
In Secs.~\ref{Sec-5:interaction-bag} and \ref{Sec-6:interaction-CQSM}
we use two models of the nucleon
to demonstrate how interactions generate a non-zero value for the
$D$-term. We use the bag model, where the interaction is provided
by the bag boundary which confines the otherwise free and 
non-interacting fermion(s). We also use the chiral quark soliton
model where the nucleon is described as a solitonic bound state 
in a strongly interacting theory of quarks, antiquarks and 
Goldstone bosons. Finally, in Sec.~\ref{Sec-7:conclusions} we 
summarize our findings and present the conclusions.

\section{Form factors of the energy-momentum tensor}
\label{Sec-2:FF-of-EMT-in-general}

The energy momentum tensor of a theory described by the 
Lagrangian ${\cal L}$ is defined by coupling the theory 
to a background gravitational field and varying the action 
$S_{\rm grav}=\int\di^4x\sqrt{-g}\,{\cal L}$ 
with respect to the background field,
\be\label{Eq:EMT-from-gravity}
	\hat T_{\mu\nu} = \frac{2}{\sqrt{-g}}\,
	\frac{\delta S_{\rm grav}}{\delta g^{\mu\nu}}\,,
\ee
where $g$ denotes the determinant of the metric. 
The matrix elements of the EMT operator in spin-$\frac12$ states
are described by three form factors \cite{Pagels} 
\ba\label{Eq:ff-of-EMT} 
    \la p^\prime| \hat T_{\mu\nu}(0) |p\rangle = \bar u(p^\prime)\biggl[
      M_2(t)\,\frac{P_\mu P_\nu}{m}
    + J(t)\,\frac{i(P_{\mu}\sigma_{\nu\rho}+P_{\nu}\sigma_{\mu\rho})\Delta^\rho}{2m}
    + D(t)\,\frac{\Delta_\mu\Delta_\nu-g_{\mu\nu}\Delta^2}{4m}\biggr]u(p)\, ,
    \ea
with states and spinors normalized by 
$\la p^\prime|p\ra = 2p^0(2\pi)^3\delta^{(3)}({\bf p}^\prime-{\bf p})$ 
and $\bar u(p) u(p)=2 m$ where $m$ denotes the mass.
We suppress spin indices for brevity, and define 
$P=(p+p')/2$, $\Delta=(p'-p)$, $t=\Delta^2$.

The form factors of the EMT in Eq.~(\ref{Eq:ff-of-EMT}) can be interpreted
\cite{Polyakov:2002yz} in analogy to the electromagnetic form factors
\cite{Sachs} in the Breit frame characterized by $\Delta^0=0$.
In this frame one can define the static EMT 
\be\label{Def:static-EMT}
    T_{\mu\nu}({\bf r},{\bf s}) =
    \int\frac{\di^3\bDelta}{(2\pi)^32E}\;\exp(i\bDelta{\bf r})\;
    \la p^\prime,S^\prime|\hat{T}_{\mu\nu}(0)|p,S\ra
\ee
with the initial and final polarization vectors of the nucleon $S$ and
$S^\prime$ defined such that they are equal to $(0,{\bf s})$ in the
respective rest-frame, where we introduce the unit vector ${\bf s}$ 
denoting the quantization axis for the nucleon spin.
The energy density $T_{00}({\bf r})$ yields the
fermion mass according to $\int\di^3r\,T_{00}({\bf r},{\bf s})=m$,
while $\varepsilon^{i j k} r_j T_{0k}^Q({\bf r},{\bf s})$ corresponds 
to the distribution of angular momentum inside the fermion.
The components of $T_{ik}({\bf r})$ constitute the stress tensor.
The form factors $M_2(t)$ and $D(t)$ are related to 
$T_{\mu\nu}({\bf r},{\bf s})$ by
\ba
    M_2(t)-\frac{t}{4m^2}\biggl(M_2(t)-2 J(t)+ D(t) \biggr)
    &=&\;\;\;\frac{1}{m}\,\int\di^3{\bf r}\, e^{-i{\bf r}\bDelta}
    \, T_{00}({\bf r},{\bf s})\,,
    \label{Eq:ff-M2}\\
    D(t)+\frac{4t}{3}\, D^\prime(t)
    +\frac{4t^2}{15}\, D^{\prime\prime}(t)
    &=& -\frac{2}{5}\,m \int\di^3{\bf r}\,e^{-i{\bf r}\bDelta}\,
    T_{ij}({\bf r})\,\left(r^i r^j-\frac{{\bf r}^2}3\,\delta^{ij}\right)\,
    , \label{Eq:ff-d1}
\ea
where the primes denote derivatives with respect to the argument. 
The explicit expressions relating 
$\varepsilon^{i j k} r_j T_{0k}^Q({\bf r},{\bf s})$ to $J(t)$,
see  \cite{Polyakov:2002yz,Lorce:2017wkb}, will not 
be needed in this work.
At zero momentum-transfer the form factors satisfy the constraints
\ba
&&   	M_2(0)
	=1, \quad
        J(0)
	=\frac12\, , \nonumber\\
&&    	D(0)=
    	-\frac{2}{5}\,m \int\di^3{\bf r}\;T_{ij}({\bf r})\,
    	\left(r^i r^j-\frac{{\bf r}^2}3\,\delta^{ij}\right)\equiv D\, .
    	\label{Eq:M2-J-d1}
\ea
The form factors $M_2(t)$ and $J(t)$ are constrained at $t=0$ because
the total energy of the fermion is equal to its mass and its spin is 1/2, 
see \cite{Lowdon:2017idv} for a recent rigorous discussion in an 
axiomatic approach. But the value of $D\equiv D(0)$ is a priori
unknown, and must be determined from experiments. 
The physical interpretation of the $D$-term is
the following. $D(t)$ is connected to the distribution of pressure and
shear forces experienced by the partons in the nucleon \cite{Polyakov:2002yz}:
$T_{ij}({\bf r})$ can be decomposed as
\be\label{Eq:T_ij-pressure-and-shear}
    T_{ij}({\bf r})
    = s(r)\left(\frac{r_ir_j}{r^2}-\frac 13\,\delta_{ij}\right)
        + p(r)\,\delta_{ij}\, . \ee
Hereby $p(r)$ describes the distribution of the ``pressure" inside the 
hadron, while $s(r)$ is related to the distribution of the ``shear forces.''
Both functions are related to each other due to the conservation of the EMT
\cite{Polyakov:2002yz}. 
Another important consequence of the EMT conservation 
is the von Laue condition \cite{von-Laue}
\be\label{Eq:stability}
    \int\limits_0^\infty \!\di r\;r^2p(r)=0 \;,
\ee
which is a necessary (but not sufficient) condition for stability.
Further worthwhile noticing properties which follow from the conservation 
of the EMT are discussed in Ref.~\cite{Goeke:2007fp}. 

\section{EMT form factors for a free Dirac particle}
\label{Sec-3:free-case}

The simplest case is the theory of a free spin $\frac12$ fermion 
described by the Lagrangian
\be\label{Eq:Lagrangian-free-case}
	{\cal L} = \bar{\Psi}(i\fslash{\partial}-m)\Psi\,.
\ee
For a free spin $\frac12$ particle Eq.~(\ref{Eq:EMT-from-gravity}) 
yields the EMT operator given by
\be\label{Eq:EMT-free}
    \hat{T}_{\mu\nu}(x) = \frac{1}{4}\,\bar\Psi(x)\,\biggl(
     i\gamma_\mu\overrightarrow{\partial}_{\!\mu}
    +i\gamma_\nu\overrightarrow{\partial}_{\!\mu}
    -i\gamma_\mu\overleftarrow{ \partial}_{\!\mu}
    -i\gamma_\nu\overleftarrow{ \partial}_{\!\mu}
    \biggr)\,\Psi(x)\;,
\ee
where the arrows indicate on which fields the derivatives act.
Evaluating the matrix elements yields
\ba
    	\la p^\prime| \hat T_{\mu\nu}(x) |p\rangle 
	= \frac{1}{4}\,\bar u(p^\prime)\biggl[
     	\,\gamma_\mu^{ } p_\nu^{ }
	+ p_\mu^{ } \gamma_\nu^{ }
	+ \gamma_\mu^{ }p^\prime_\nu
	+p^\prime_\mu \gamma_\nu 
	\,\biggr]u(p)\, e^{i(p^\prime-p)x}\,. \label{Eq:ff-of-EMT-free-1} 
\ea
Exploring the Gordon identity we can rewrite this result as
\be
    \la p^\prime| \hat T_{\mu\nu}(x) |p\rangle = \bar u(p^\prime)\biggl[
	\frac{P_\mu P_\nu}{m} + \frac12\;
	\frac{i(P_{\mu}\sigma_{\nu\rho}+P_{\nu}\sigma_{\mu\rho})\Delta^\rho}{2m}
	\biggr]u(p)\, e^{i(p^\prime-p)x}\,,
    \label{Eq:ff-of-EMT-free-2} 
\ee
from which we read off the predictions of the free Dirac theory for the 
EMT form factors, namely
\be\label{Eq:ff-of-EMT-free-3}
	M_2(t) = 1	\, , \;\;\;
	J(t) = \frac12	\, , \;\;\;
	D(t) = 0 	\, .
\ee
Several comments are in order. The form factors are constant functions
of $t$ as expected for a free point-like particle, and we consistently
recover the general constraints for $M_2(t)$ and $J(t)$ at $t=0$ in 
Eq.~(\ref{Eq:M2-J-d1}). The value of the $D$-term is therefore the 
only non-trivial result from this exercise: it is remarkable 
it vanishes for a free point-like fermion \cite{Footnote-Majorana}.
 
It is important to remark that the vanishing of the $D$-term in the free 
case was implicitly known in literature, see e.g.\ \cite{Donoghue:2001qc}
where quantum corrections to the metric were studied. 
Although a quantum gravity theory is not yet known, the leading 
quantum corrections can be computed from the known low energy structure 
of the theory \cite{Donoghue:1993eb}. These calculations are challenging
\cite{Khriplovich:2002bt,BjerrumBohr:2002kt,Khriplovich:2004cx}.
But the ``tree level'' results for EMT form factors were obtained
unambiguously already in \cite{Donoghue:2001qc}. Our free field calculation, 
Eq.~(\ref{Eq:ff-of-EMT-free-3}), agrees with Ref.~\cite{Donoghue:2001qc}.
The loop corrections to the Reissner-Nordstr\"om and Kerr-Newman metrics
\cite{Khriplovich:2002bt,BjerrumBohr:2002kt,Khriplovich:2004cx}
show how (QED, gravity) interactions generate quantum long-range 
contributions to the stress tensor. A consistent description of the 
$D$-term requires, however, the full picture of the stress tensor
including short-distance contributions which cancel exactly the 
long-distance ones in Eq.~(\ref{Eq:stability}). The results of
these works therefore do not allow us to gain insights on how
much these corrections contribute to the $D$-terms of
elementary (and charged) fermions.

\section{\boldmath Heuristic consistency argument
	why $\bm{D=0}$ for a free fermion}
\label{Sec-4:heuristic}

The vanishing of the $D$-term of a free fermion can 
be made plausible on the basis of a heuristic argument
which was already helpful in discussing the EMT densities 
in the bosonic case \cite{Hudson-PS-I}.
The argument explores the 3D-density framework which strictly 
speaking requires the particle to be heavy such that relativistic 
corrections can be neglected. 
The argument is based on two assumptions:
(i) form factors are $t$-independent constants in the free theory case, and 
(ii) energy density is formally 
given by $T_{00}(\vec{r}\,)= m\;\delta^{(3)}(\vec{r}\,)$
for a heavy particle \cite{Footnote-heavy-mass}.

Per assumption (i) we can replace the form factors in Eq.~(\ref{Eq:ff-M2}) 
by their values at zero-momentum transfer. Next, we notice that the result 
in the square brackets in the following equation must be zero to comply with 
assumption (ii),
\be
	\frac{1}{m}\,\int\di^3{\bf r}\, e^{-i\bDelta{\bf r}}
    	\, T_{00}({\bf r}) =
	M_2(0)-\frac{t}{4m^2}
	\underbrace{\left[M_2(0)-2 J(0)+D(0) \right]}_{
		\displaystyle\stackrel{!!}{=} 0}
	\stackrel{!} = 1 	\, .
    	\label{Eq:ff-M2a}
\ee
With the constraints in Eq.~(\ref{Eq:M2-J-d1}) it is clear that
$M_2(0)-2 J(0)=0$. From this it then immediately follows that the 
$D$-term must vanish for a point-like particle for consistency reasons. 

This is nothing but a heuristic argument. But it is nevertheless helpful 
to make it plausible why the $D$-term of a free fermion vanishes. 
From this argument it is also clear why in the interacting case one may in 
general encounter a non-zero $D$-term: when interactions are present form 
factors can no longer be expected to be $t$-independent constants, and 
$D(t)$ in general do not need to be zero. An extended internal structure 
implies a non-zero $D$-term along the same lines: now
$T_{00}(\vec{r}\,)\neq m\;\delta^{(3)}(\vec{r}\,)$ 
and form factors exhibit a generic $t$-dependence, e.g, 
$M_2(t)=1+\frac16\,\la r^2\ra \,t + {\cal O}(t^2)$
\cite{Goeke:2007fp}. 

In App.~\ref{App-A} we include another heuristic argument why 
the $D$-term vanishes for elementary fermions but not for elementary
bosons, based on a simple analysis of the structure of the Lagrangians.

\newpage
\section{\boldmath Emergence of the $D$-term from bag boundary forces}
\label{Sec-5:interaction-bag}

The bag model describes one or several non-interacting fermions
confined inside a ``bag'' which, in its rest frame, is a spherical 
region of radius $R$ carrying the energy density $B>0$.
If $N_c=3$ quarks or a $\bar qq$-pair are placed inside the bag in a 
color-singlet state, this yields the popular model of hadrons with 
confinement simulated by the bag boundary condition \cite{Chodos:1974je}. 
The Lagrangian of the bag model can be written as \cite{Thomas:2001kw}
\be\label{Eq:Lagrangian-bag}
   {\cal L} = \biggl(\bar\psi\, (i\fslash{\partial}-m)\psi-B\biggr)\,\Theta_V
            + \frac12\,\bar\psi\,\psi\:\eta^\mu\partial_\mu\Theta_V \;,
\ee
where
$\Theta_V=\Theta(R-r)$, 
$\delta_S=\delta(R-r)$,
$\eta^\mu=(0,\vec{e}_r)$, 
$\vec{e}_r = \vec{x}/r$,
$r=|\vec{x}\,|$ in the bag rest frame.
The indices $V$ and $S$ denote respectively the volume and the surface of 
the bag. The boundary condition for the fields is equivalent to the
statement that there is no energy-momentum flow out of the bag, i.e.\ 
$\eta_\mu T^{\mu\nu}(t,\vec{r})=0$ for $\vec{r}\in S$.

The starting point is as follows.
If no bag boundary condition is present, i.e.\ in the limit $R\to\infty$
in Eq.~(\ref{Eq:Lagrangian-bag}), we deal with the free Lagrangian 
(\ref{Eq:Lagrangian-free-case}) with an additive constant $B$ which is 
irrelevant and can be discarded. In such a free theory the $D$-term is 
zero, as we have shown in Sec.~\ref{Sec-3:free-case}.

Next let us discuss what happens if we solve the theory with the bag 
radius $R$ kept finite. This means we effectively introduce an 
interaction acting on the otherwise free fermion. We will see that now
a non-zero $D$-term emerges. Below we quote only the main steps 
needed in our context. The details of this calculation will be reported 
elsewhere \cite{Neubelt-et-al}. 

The equations of motion of the theory (\ref{Eq:Lagrangian-bag})
are $(i\fslash{\partial}-m)\psi = 0$ for $r<R$, while at
the surface $\vec{x}\in S$ the linear boundary condition
$i\fslash{\eta}\,\psi = \psi$ and the non-linear boundary condition
$\frac12\,\eta_\mu\partial^\mu(\bar\psi\psi) =- B$ hold. 
The ground state solution has positive parity and is 
given by the wave-function 
\be\label{Eq:bag-wave-function}
    \psi(t,\vec{x}) = e^{-i\omega t/R} \;\frac{A}{\sqrt{4\pi}}\,
    \left(\begin{array}{l}
        \alpha_+j_0(\omega \, r/R)\,\chi_s \phantom{\displaystyle\frac11}\\
        \alpha_-j_1(\omega \, r/R)\,i\vec{\sigma}\vec{e}_r\chi_s 
    \end{array}\right)
	\, , \;\;\;
\ee
where $\alpha_\pm=\sqrt{1\pm mR/\Omega}$ and $\Omega=\sqrt{\omega ^2+m^2R^2}$,
$\omega $ denotes the lowest solution of the equation
$\omega = (1-mR-\Omega)\,\tan\omega$,
$\sigma^i$ are Pauli matrices, $\chi_s$ are two-component spinors.
The normalization 
$\int\di^3x\;\psi^\dag(\vec{x},t)\,\psi(\vec{x},t) = 1$
fixes the constant $A$.
If $N_c$ fermions are placed in the bag the $D$-term is given by 
\be\label{Eq:D-term-bag}
	D= \frac13\,M\,N_c\,\frac{A^2 R^4\!}{\omega^4}\;\alpha_+\alpha_-
	\biggl(
	-\frac{\omega ^3}{3}
	+\frac{5}{4}\,(\omega-\sin\omega \,\cos\omega)
	-\frac{\omega }{2}\,\sin^2\omega 
\biggr)\,.
\ee
where $M=N_c\Omega/R+\frac43\,\pi\,B\,R^3$ is the mass of the system. 
One can show that always $D<0$ in this model \cite{Neubelt-et-al}. 
For $N_c=3$ colors and assuming the fermions to be massless quarks
(in which case $\omega=2.04\dots$) one obtains $D=-1.145$ in
agreement with the numerical bag model calculation of nucleon GPDs 
and EMT form factors from Ref.~\cite{Ji:1997gm}. 

As an application of Eq.~(\ref{Eq:D-term-bag}) it is insightful to consider
the limit $mR\to\infty$ where $\omega\to\pi$, and the $D$-term becomes
\be\label{Eq:D-term-bag-non-rel}
	D = N_c^2\,\frac{(-4\,\pi^2+15)}{45} \,.
\ee
This result can be interpreted in two ways.

For the first interpretation we may assume that $m$ is fixed and $R$ 
becomes much larger \cite{Footnote:limit-R} 
than the Compton wave length of the particle, $R\gg 1/m$. This means 
the ``interaction'' decreases, as the confined particle(s) can 
occupy an increasing volume with the boundary being ``moved'' further 
and further away. However, no matter how far away we move the boundary
\cite{Footnote:limit-R}:
{\it some} interaction remains, and generates a non-zero $D$-term.

For the second interpretation we may assume a fixed $R$ and $m\to\infty$.
This is known as the non-relativistic limit, in which $\alpha_-\to 0$
and the lower component of the spinor in (\ref{Eq:bag-wave-function}) 
vanishes. The $D$-term in Eq.~(\ref{Eq:D-term-bag})
is proportional to $\alpha_-$ which vanishes, and to the mass of the system 
which behaves as $M\to N_cm$ for $m\to\infty$. The product $M\alpha_-$ is
finite in the limit $m\to\infty$. As a result the $D$-term assumes a 
finite value as quoted in Eq.~(\ref{Eq:D-term-bag-non-rel}). 
This result demonstrates that also non-relativistic systems have 
a $D$-term, i.e.\ this property is not a relativistic effect.
For a detailed discussion of the $D$-term in the bag model 
we refer to \cite{Neubelt-et-al}.

One virtue of the bag model is its transparency, which we 
explored here to learn insightful lessons about the $D$-term. 
One caveat is that it does not comply with chiral symmetry 
which is incorporated in the model discussed next.

\newpage
\section{Chiral interactions and the $\bm{D}$-term of nucleon}
\label{Sec-6:interaction-CQSM}

The spontaneous breaking of chiral symmetry is the dominant feature
of strong interactions in the non-perturbative low-energy regime.
A theoretically consistent and phenomenologically successful 
model of baryons based on chiral symmetry breaking is 
the chiral quark-soliton model \cite{Diakonov:1987ty} defined in 
the SU(2) flavor-sector by \cite{Diakonov:1984tw,Dhar:1985gh}
\be\label{Eq:CQSM}
    {\cal L}_{\rm eff}=\bar{\Psi}\,(i\fslash{\partial\,}-M\,U^{\gamma_5})\Psi\,,
    \;\;\; U^{\gamma_5}=\exp(i\gamma_5\tau^a\pi^a/F_\pi)
\ee
where $F_\pi=93\,{\rm MeV}$ denotes the pion decay constant.
Besides the emergence of Goldstone bosons, another consequence 
of chiral symmetry breaking is the dynamically generated 
``constituent'' quark mass $M\simeq 350\,{\rm MeV}$.
The effective theory (\ref{Eq:CQSM}) was derived from the instanton 
model of the QCD vacuum \cite{Diakonov:1983hh,Diakonov:1985eg} which 
provides a microscopic picture of the dynamical breaking of chiral 
symmetry, see  \cite{Diakonov:2000pa} for reviews.

In order to solve the strongly coupled theory (\ref{Eq:CQSM})
(the coupling constant of quark and pion field is $M/F_\pi\sim 3.8$)
a non-perturbative method based on the limit of a large number of 
colors $N_c$ is used.
In this limit the functional integration over $U$-fields in 
Eq.~(\ref{Eq:CQSM}) is solved in the saddle-point approximation 
by evaluating the model expressions at the static solitonic field 
$U(\vec{x})$ and integrating over the zero-modes of the soliton 
solution. The spectrum of the Hamiltonian of the effective theory 
(\ref{Eq:CQSM}), $H=-i\gamma^0\gamma^k\partial_k+M\gamma^0U^{\gamma_5}(\vec{x})$,
contains continua of positive energies $E>M$ and negative energies $E<-M$,
and a discrete level with an energy $-M < E_{\rm lev} < M$. The nucleon
state is obtained by occupying the discrete level and the states of
negative continuum and subtracting the free negative continuum 
(``vacuum subtraction'').
The solitonic field $U(\vec{x})$ is determined from a self-consistent
variational procedure which minimizes the soliton energy.
In the physical situation the soliton size is
$R_{\rm sol} \sim M^{-1}$ \cite{Diakonov:1987ty}. 

GPDs and EMT form factors including the $D$-term were studied 
in this model \cite{Petrov:1998kf,Schweitzer:2002nm,
Ossmann:2004bp,Wakamatsu:2007uc,Goeke:2007fp,Goeke:2007fq}. 
As a demonstration of the consistency of this effective chiral
theory let us mention that in this model the GPDs satisfy 
polynomiality \cite{Schweitzer:2002nm}, the Ji sum
rule is valid \cite{Ossmann:2004bp}, the von Laue condition holds, 
the model correctly reproduces the leading non-analytic 
terms of the EMT form factors \cite{Goeke:2007fp}, and
agrees with available lattice QCD data \cite{Goeke:2007fq}. 

We will now show that the $D$-term vanishes when one ``switches off''
the chiral interactions in this model. This can be formally done by 
replacing $U\to 1$ in Eq.~(\ref{Eq:CQSM}) which yields the free theory.
One way to practically implement this limit is to consider the formal
limit $M\,R_{\rm sol}\to\infty$. As the soliton size increases 
the discrete level energy decreases and approaches the negative 
continuum \cite{Diakonov:1987ty}. Since in this limit the spatial 
extension of the solitonic field $U(\vec{x})$ grows, its gradients 
$\nabla U(\vec{x})$ decrease. This allows one to expand model
expressions in terms of gradients of the chiral field.

The expression for the $D$-term valid in such a large soliton expansion
was derived in \cite{Schweitzer:2002nm} and is given by
\be\label{Eq:d1-gradient-expansion}
	D = -\,F_\pi^2 M_N\int\di^3x\,P_2(\cos\vartheta)\,
	\vec{x}^{\,2}\,{\rm tr}_F[\nabla^3 U][\nabla^3 U^\dag]+\dots
\ee
where ${\rm tr}_F$ is the trace over flavor indices, $M_N$ denotes
the nucleon mass, and the dots indicate higher order derivatives. 
Notice that the expression (\ref{Eq:d1-gradient-expansion}) is 
quasi model-independent: it is the leading contribution in the 
chiral expansion of the $D$-term from which one can derive the 
leading non-analytic terms \cite{Goeke:2007fq}. The second 
Legendre polynomial reflects that the $D$-term is related 
to the traceless part of the stress tensor \cite{Polyakov:2002yz}. 

After these preparations we can now discuss what happens in 
the formal limit when we ``switch off'' the chiral interactions 
and $U\to 1$. In this limit all gradients vanish in 
Eq.~(\ref{Eq:d1-gradient-expansion}) and we recover that $D=0$
which is the free field theory prediction obtained in 
Eq.~(\ref{Eq:ff-of-EMT-free-3}).
This shows that the $D$-term in the chiral quark soliton model 
is due to the chiral interactions which define and characterize
this model.  

Let us stress that the above discussion applies only to the formal
limit $U\to 1$ which we implemented by means of the large soliton 
expansion. Only in this limit it is justified to expand model expressions
in powers of the derivatives of the chiral field. 
In the physical situation the soliton size is such that $M\,R_{\rm sol}\sim 1$
and no such expansion is allowed (though it can be used to derive 
chiral leading non-analytic contributions, and it may give useful rough 
estimates). In order to obtain 
in the physical situation reliable model predictions for the 
$D$-term, and a pressure satisfying the von Laue condition 
(\ref{Eq:stability}), it is necessary to evaluate numerically 
the full model expression \cite{Goeke:2007fp}.

\newpage
\section{Conclusions}
\label{Sec-7:conclusions}

The $D$-term of a free non-interacting fermion vanishes. This is
a simple prediction of the free Dirac equation which is, in principle,
analog to the prediction $g=2$ for the anomalous magnetic moment of a
charged point-like fermion. This result is remarkable for several reasons
and has interesting implications.

The prediction of a vanishing $D$-term from the free Dirac equation
should be contrasted with the bosonic case. The free Klein-Gordon
equation predicts an intrinsic non-zero $D$-term already for 
free and non-interacting bosons.
When interactions are introduced in bosonic theories, the 
value of $D$ is in general affected and, depending on 
the theory, the effect can be sizable \cite{Hudson-PS-I}. 
However, in the fermionic case interactions do not modify the 
$D$-term, but {\it generate} it. In other words, the $D$-term
of a spin-$\frac12$ particle is entirely of dynamical origin.

We have provided an heuristic consistency argument which
makes it plausible why the $D$-term of a free point-like
spin $\frac12$ particle should vanish. While not a rigorous 
derivation, this argument was already successfully applied 
to explain why a free point-like boson must have $D=-1$ \cite{Hudson-PS-I}. 

We have explored two dynamical models of the nucleon to 
illustrate how the $D$-term is generated in interacting systems.
In the bag model we have shown how a non-zero $D$-term emerges 
when we ``switch on'' interactions which in this model are formulated
in terms of boundary conditions which confine otherwise free fermions.
We used also the chiral quark soliton model where we have shown how
the $D$-term vanishes when the strongly coupled chiral interactions 
in that model are ``switched off.'' 
These are simple models of the nucleon, but these results solidify
our conclusions: in a fermionic system the $D$-term is generated
by dynamics, it arises entirely from interactions.

The calculations of the nucleon $D$-term in models, lattice QCD, 
or dispersion relations 
\cite{Ji:1997gm,Petrov:1998kf,Schweitzer:2002nm,Ossmann:2004bp,
Wakamatsu:2007uc,Goeke:2007fp,Goeke:2007fq,Cebulla:2007ei,Kim:2012ts,
Hagler:2003jd,Gockeler:2003jf,Negele:2004iu,Bratt:2010jn,Pasquini:2014vua}
give therefore insights which are completely due to the underlying 
(effective, model, chiral, QCD) dynamics. 
With its relation to the internal forces and the stress tensor
\cite{Polyakov:2002yz} the $D$-term emerges therefore as a valuable 
window to gain new insights on the structure of composite particles,
and especially the QCD dynamics inside the nucleon.

In any case, all presently known matter is composed of what we 
consider elementary fermions, which indicates the importance to 
study this interesting particle property in more detail.
Knowledge of EMT form-factors can be applied to the spectroscopy 
of the hidden-charm pentaquarks observed at LHCb 
\cite{Eides:2015dtr,Perevalova:2016dln}.
Also the EMT form factors of mesons can be inferred from data
and this information may help to discriminate usual from exotic
\cite{Polyakov:1998ze,Kawamura:2013wfa}. 

It will be very exciting to learn about the $D$-term from
lattice QCD calculations and experiment and the perspectives
are good. 
After first, vague and model-dependent glimpses on the nucleon
$D$-term from the HERMES experiment \cite{Ellinghaus:2002bq} 
one may expect more quantitative insights from experiments 
at Jefferson Lab \cite{JLab,Jo:2015ema}, COMPASS at CERN
\cite{Joerg:2016hhs}, and the envisioned future 
Electron-Ion-Collider \cite{Accardi:2012qut}.

\

\

\noindent{\bf Acknowledgments.}
We would like to thank C\'edric Lorc\'e and Maxim Polyakov for
helpful discussions.
This work was supported in part by the National Science Foundation 
(Contract No.\ 1406298).

\appendix

\section{\boldmath Why can the Klein-Gordon equation give $D\neq 0$
	but Dirac equation cannot?}
\label{App-A}

One may wonder why the Klein-Gordon equation can naturally predict a 
non-zero $D$-term, but the Dirac equation cannot. It is instructive 
to review how this happens. 
The $D$-term appears in the decomposition of the matrix elements 
of the EMT operator (\ref{Eq:ff-of-EMT}) with the same structure 
$\Delta_\mu\Delta_\nu-g_{\mu\nu}\Delta^2$ in the bosonic and fermionic
case. In spin-zero case such a 
structure emerges already from the kinetic term in the Lagrangian 
${\cal L}=\partial_\mu\Phi^\ast\partial^\mu\Phi-V$. The kinetic term 
contains two field derivatives and generates the contribution 
$\partial^\mu\Phi^\ast\,\partial^\nu\Phi+\partial^\mu\Phi\,\partial^\nu\Phi^\ast$
to the EMT operator. This is sufficient to generate the needed structure 
$\Delta_\mu\Delta_\nu-g_{\mu\nu}\Delta^2$ in the EMT matrix elements
even in the absence of interactions, when  $V = m^2\Phi^\ast\Phi$ in 
the free case. Interactions may affect the $D$-term (and make its value 
more or less negative, but preserving its sign in all theories studied so far).
The main point is, however, that even in the free theory a non-zero 
$D$-term arises in the spin-zero case \cite{Hudson-PS-I} and this can 
be naturally traced back to the Lagrangian containing 2 derivatives of 
the fields needed to generate in the EMT a structure proportional
to $\Delta_\mu\Delta_\nu-g_{\mu\nu}\Delta^2$.
In contrast to this, in the case of free Dirac fields the Lagrangian 
contains only one derivative, and consequently no $D$-term can be 
generated. Let us notice that {\it if} interactions are present they
of course may generate a $D$-term in the fermionic case, see 
sections \ref{Sec-5:interaction-bag} and \ref{Sec-6:interaction-CQSM}
for some illustrations.

\newpage


\begin{thebibliography}{99}
\bibitem{Footnote-Majorana}
	This result is equally valid for  Dirac or Majorana fermions, and 
	Weyl fermions with appropriate modifications to the normalization 
	of the spinors and field operators.
\bibitem{Footnote-heavy-mass}
	In order to define the heavy 
	mass limit it is strictly speaking necessary to introduce an 
	additional scale in the problem, so we can tell with respect 
	to what the mass $m$ is supposed to be large. No such scale
	is available in the free theory case. Our assumption that 
	$T_{00}(\vec{r}\,)= m\;\delta^{(3)}(\vec{r}\,)$ is therefore 
	of formal character. See Ref.~\cite{Hudson-PS-I} for a 
	detailed discussion.
\bibitem{Footnote:limit-R}
	At this point $R$ can be assumed arbitrarily large (e.g.\ 
	``radius of the visible universe'') but must be finite, 
	as otherwise the implementation of the boundary conditions 
	would not have been justified. The strict limit $R\to\infty$ 
	(which reproduces a free spin $\frac12$ particle with a $D$-term 
	which is zero) must be taken on the level of the Lagrangian 
	(\ref{Eq:Lagrangian-bag}) before starting the bound state calculation, 
	as discussed above.
\end{thebibliography}
\end{document}